\documentclass[aps,pra,twocolumn,amsmath,showpacs,endfloats]{revtex4}
\usepackage{graphicx}

\begin{document}
\title{Light scattering from self-affine fractal silver
surfaces with nanoscale cutoff: Far-field and near-field calculations}

\author{Jos\'e A. S{\'a}nchez-Gil}
\email{j.sanchez@iem.cfmac.csic.es}
\homepage{http://www.iem.cfmac.csic.es/departamentos/espvib_procmult/jsanchez.htm}
\author{Jos\'e V. Garc{\'\i}a-Ramos}
\affiliation{Instituto de Estructura de la Materia,
Consejo Superior de Investigaciones Cient{\'\i}ficas,
Serrano 121, E-28006 Madrid, Spain
}
\author{Eugenio R. M{\'e}ndez}
\affiliation{Divisi{\'o}n de F{\'\i}sica Aplicada, Centro de
Investigaci{\'o}n Cient{\'\i}fica y de Educaci{\'o}n Superior de
Ensenada \\ Ensenada, Baja California 22800, M{\'e}xico}
\date{\today}

\begin{abstract}
We study the light scattered from randomly rough, one-dimensional
self-affine fractal silver surfaces with nanoscale lower cutoff,
illuminated by $s$- or $p$-polarized Gaussian beams a few microns
wide. By means of rigorous numerical calculations based on the Green
theorem integral equation formulation, we obtain both the far- and
near-field scattered intensities. The influence of diminishing the
fractal lower scale cutoff (from below a hundred, down to a few
nanometers) is analyzed in the case of both single realizations and
ensemble average magnitudes. For $s$ polarization, variations are
small in the far field, being only significant in the higher spatial
frequency components of evanescent character in the near field. In the
case of $p$ polarization, however, the nanoscale cutoff has remarkable
effects stemming from the roughness-induced excitation of
surface-plasmon polaritons. In the far field, the effect is noticed
both in the speckle pattern variation and in the decrease of the total
reflected energy upon ensemble averaging, due to increased
absorption. In the near field, more efficient excitation of localized
optical modes is achieved with smaller cutoff, which in turn leads to
huge surface electric field enhancements.
\end{abstract}
\maketitle

\section{INTRODUCTION}
\label{sec_int}

Since the early days of fractality,\cite{mandel} the scattering of
electromagnetic (EM) waves from fractal surfaces has been a field
of intense activity. Physical  fractals appear ubiquitously
in nature, possessing fractal properties within a broad, however
finite, range of scales. Therefore the study of
classical wave scattering from fractals is a problem of interest not
only from a fundamental point of view, but also from the practical
knowledge that  probing technologies such as surface optical
characterization, remote sensing, radar, sonar,  etc. can yield
about a wide variety of systems. In fact, it is now well understood
that many naturally occurring surfaces exhibit scale invariance,
particularly in the form of self-affinity.\cite{mandel,feder,bara}

There exists a large amount of theoretical works devoted to wave
scattering from fractal surfaces.\cite{berry79,jake83,jagg87,jull89,%
sinha89,west90,jagg90,yang93,mcsha95,lin95,sava,chen96,shepp96,%
ao97,wrm97,simon98,zhao98,frances00,simon}  Most of them make use of
approximations, such as Kirchhoff approximation (KA) or perturbation
methods, in order to obtain analytical expressions that are useful in
certain regimes, but thereby imposing a constraint on the length
scales over which fractality might be present.  Others include
calculations within the Rayleigh hypothesis and/or restricted to
perfectly conducting surfaces. And only very recent works are capable
of dealing with arbitrarily rough metal or dielectric surfaces,%
\cite{ao97,wrm97,frances00,simon} on the basis on the Green's
theorem integral equation formulation for rough surface scattering.%
\cite{ann,nieto,josaa91}

On the other hand, in addition to far field scattering, the near
EM field on self-affine fractals has attracted a great deal of
attention.\cite{sharep,sha96,oc97,jcp98,poli98,gres99,prb00} To a
large extent, the interest lies in the observation in the near field
intensity distributions, through photon scanning tunneling microscopy
(PSTM), of large concentrations of EM field intensity on {\it bright
spots}).\cite{gres99,tsai94,bozh95,bozh96,zhang98,bozh98,mark99} The
appearance of such {\it localized optical modes} is mediated by the
roughness-induced excitation of surface-plasmon polaritons (SPP) on
very rough surfaces with subwavelength features.  The occurrence of
large surface EM fields is indeed crucial to the EM mechanism in
Surface-Enhanced Raman Scattering (SERS) and to other non-linear
surface optical processes.\cite{sharep,jcp98,poli98,gres99,prb00,%
mark99,shg} Furthermore, the predicted, enormously high values
of the surface electric field intensity have been related to the
recent observation of SERS single molecule 
probing.\cite{nie97,kneipp97,xu99} 

It is our aim to employ the above mentioned rigorous Green's theorem
formulation to study the far-field and near-field scattered from
one-dimensionally rough, self-affine fractal Ag surfaces, restricted
to one dimension for the sake of computational limitations.
The scattering model is described in Sec.~\ref{sec_eq}. We focus
on fractals preserving self-affinity from tens of microns to
subwavelength dimensions. Actually, we analyze the influence of the
lower scale cutoff as is reduced from $\sim 100$ nm (as used in Refs.
\onlinecite{ao97,wrm97,oc97,jcp98,prb00}) to a few nanometers. It should be
mentioned that the effect on the far field of the lower scale cutoff
in deterministic Koch fractals has been previously studied.\cite{ao97}
The results for the far-field and near-field calculations are
presented in Secs.~\ref{sec_ff} and~\ref{sec_nf}, respectively.
Finally, the main conclusions derived from this work
are summarized in Sec.~\ref{sec_con}.

\section{SCATTERING MODEL}
\label{sec_eq}

\begin{figure}[h]
\includegraphics[width=.9\columnwidth]{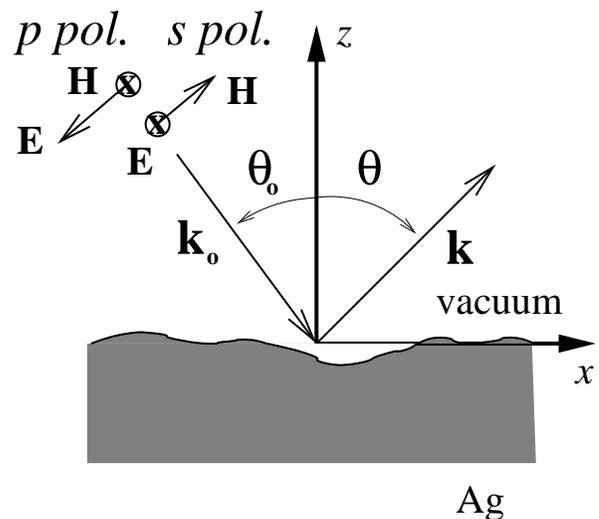}
\caption{Illustration of the scattering geometry.}
\label{fig_sca}
\end{figure}

The scattering geometry is depicted in Fig.~\ref{fig_sca}.  A
monochromatic, linearly polarized Gaussian beam of frequency $\omega$
and $1/e$ amplitude half width $W$ impinges at an angle $\theta_0$ on
a randomly rough surface $z=\zeta(x)$. The rough interface separates
vacuum from a semi-infinite silver volume occupying the lower
half-space [$z<\zeta(x)$] and characterized by an isotropic,
frequency-dependent homogeneous dielectric function
$\epsilon(\omega)$.

With the aim of solving this scattering problem for arbitrarily large
roughness parameters, we make use of the scattering integral equations
based on the the application of Green's second integral
theorem.\cite{nieto} It is well known that this full vectorial
formulation is considerably simplified when restricted to 1D surfaces
and linear polarization, thereby being reduced to four scalar integral
equations for the only nonzero component of either the electric field
amplitude
\[
  {\bf E}^{(s)}\equiv E^{(s)}({\bf r},\omega){\bf \hat{y}},
\]
with
\[
  {\bf H}^{(s)}\equiv H^{(s)}_x({\bf r},\omega){\bf \hat{x}} +
                          H^{(s)}_z({\bf r},\omega){\bf \hat{z}},
\]
for $s$ polarization, or the magnetic field amplitude
\[
  {\bf H}^{(p)}\equiv H^{(p)}({\bf r},\omega){\bf \hat{y}},
\]
with
\[
  {\bf E}^{(p)}\equiv E^{(p)}_x({\bf r},\omega){\bf \hat{x}} +
                          E^{(p)}_z({\bf r},\omega){\bf \hat{z}},
\]
for $p$ polarization.\cite{ann,josaa91} Two of these surface integral
equations can be used as extended boundary conditions, which
(recalling the continuity relations across the interface) lead to a
system of two coupled integral equations for the surface field and its
normal derivative. This can be solved numerically in the form of
linear equations by means of a quadrature scheme.\cite{ann,josaa91}
Once these source functions are obtained, it is straightforward to
calculate the far field scattered intensity,\cite{josaa91} and in
general the electric (magnetic) field amplitude for $s$ ($p$)
polarization at any point in vacuum or inside the metal from two of
the integral equations.\cite{jcp98} In addition, from the latter
integral equations, the corresponding expressions for the magnetic
(electric) field amplitudes for $s$ ($p$) polarization can be simply
obtained by making use of Maxwell equations (cf. Ref. \onlinecite{prb00} in
the case of $p$ polarization). These expressions can be useful in near
electric or magnetic field calculations. However, they exhibit
non-integrable singularities when approaching the surface profile.

In this regard, the calculation of the surface magnetic field
intensity for $s$ polarization has been done in a simple manner by
exploiting the connection between its normal and tangential components
on the vacuum side
\begin{subequations}\begin{eqnarray}
 H^{(s)}_n(x)& =& -\frac{\imath c}{\omega} \gamma^{-1}
 \frac{dE^{(s)}(x)}{dx}, \\
 H^{(s)}_t(x)& =& \frac{\imath c}{\omega} \gamma^{-1} F^{(s)}(x),
\end{eqnarray}\label{eq_smf_s}\end{subequations}
with  the surface electric
field and its normal derivative from the vacuum side,
\begin{subequations}\begin{eqnarray}
 E^{(s)}(x)&=&E^{(s,>)}(x,\zeta(x)), \\
 F^{(s)}(x)& =& \gamma \left.
\frac{\partial E^{(s,>)}({\bf r})}{\partial n}\right| _{z=\zeta^{(+)}(x)},
\end{eqnarray}\label{eq_sef_s}\end{subequations}
with $\gamma\equiv [1+[\zeta'(x)]^2]^{1/2}$.
The corresponding expressions for the surface electric field components
for $p$ polarization are given by:\cite{prb00}
\begin{subequations}\begin{eqnarray}
 E^{(p)}_n(x)& =& \frac{\imath c}{\omega} \gamma^{-1}
   \frac{dH^{(p)}(x)}{dx}, \\
 E^{(p)}_t(x)& =& -\frac{\imath c}{\omega} \gamma^{-1}
L^{(p)}(x),
\end{eqnarray}\label{eq_sef_p}\end{subequations}
where
\begin{subequations}\begin{eqnarray}
 H^{(p)}(x)&= &H^{(p,>)}(x,\zeta(x)) \\
 L^{(p)}(x)& =&\gamma \left.
 \frac{\partial H^{(p,>)}({\bf r})}{\partial n}\right| _{z=\zeta^{(+)}(x)}.
\end{eqnarray}\label{eq_smf_p}\end{subequations}
As mentioned above, $E^{(s)}(x),F^{(s)}(x),H^{(p)}(x),L^{(p)}(x)$
constitute the source functions of the
integral equations in this scattering configuration.

As a model describing many naturally occurring surface growth
phenomena exhibiting self-affine fractality, we have chosen that given
by the trace of a fractional Brownian motion through Voss'
algorithm,\cite{feder,voss88,voss89} In addition, this model
yields self-affine fractal structures that resemble fairly well the
properties of some SERS metal substrates.\cite{bozh96,zhang98,mark99}
The ensembles of realizations thus generated are characterized by
their fractal dimension $D=2-{\cal H}$ (${\cal H}$ being the Hurst
exponent) and rms height $\delta$. (In order to avoid the inherent
ambiguity in the definition of the rms height for self-affine
fractals, $\delta$ refers in our calculations to the rms height
defined over the entire fractal profile with length
$L_f=51.4\,\mu$m. Recall that $\delta$ depends on the length $\Delta
x$ over which it is measured through\cite{simon} $\delta=l^{1-{\cal
H}}\Delta x^{\cal H}$, $l$ being the topothesy. Thus the topothesy $l$
will be also given. From each generated fractal profile with $N_f$
points and length $L_f$, sequences of $N$ points (with constant
$N/N_f$) are extracted to obtain similar profiles with identical
properties except for the lower scale cutoff as determined by
$\xi_L=L_f/N_f$. (Strictly speaking, $\xi_L$ should be given by the
minimum length scale above which the ensemble of such realizations
exhibit self-affinity, resulting in a value typically larger than the
mere discretization cutoff.\cite{feder,jcp98})

From now on, we will focus below on the fractal dimension $D=1.9$.
This choice is justified for we expect the influence of decreasing
$\xi_L$ to be more significant for larger fractal dimensions. And
recall that the effect of varying $D$ has been already studied on the
far field \cite{wrm97} and the near field \cite{jcp98}, albeit for a
relatively large lower cutoff $\xi_L$. The values of the lower scale
cutoffs hitherto considered are $\xi_L=51.4, 25.7, 12.85,$ and 6.425
nm, resulting from sequences of $N=102, 205, 410,$ and 819 points
extracted from profiles with $N_f=1024, 2048, 4096,$ and 8192
points. The length of all realizations is thus
$L=L_f(N/N_f)=5.14\,\mu$m.  Actually, the final number of sampling
points per realization used in the numerical calculations is
significantly higher for the sake of accuracy: $N_p=n_i N$ as obtained
by introducing $n_i=$4-10 cubic-splined interpolating points, the
latter being chosen on the basis of numerical convergence tests.

\section{FAR FIELD}
\label{sec_ff}

\begin{figure}[h]
\includegraphics[width=\columnwidth]{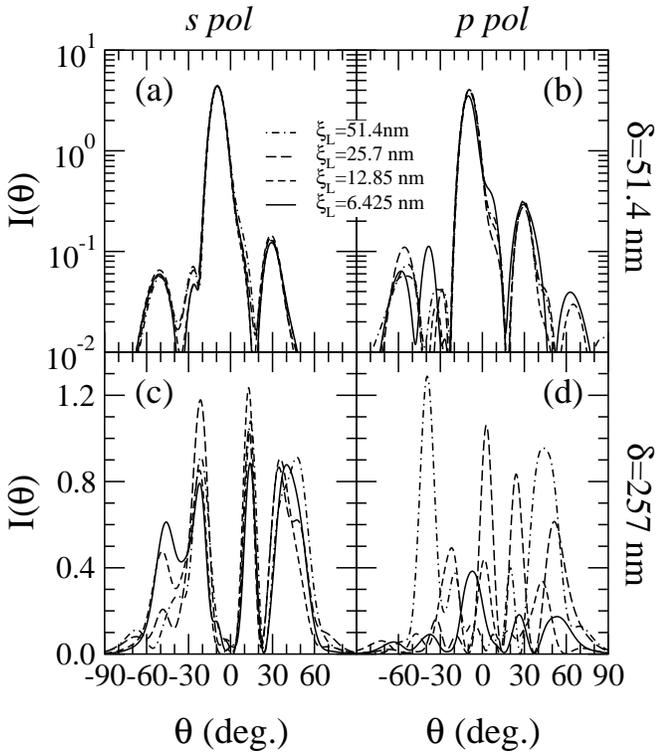}
\caption{Angular distribution of the speckle pattern intensity scattered
        from Ag fractal surface profiles with $L=5.14\,\mu$m, $D=1.9$, and
        $\delta=51.4$ nm (top, a-b) and $\delta=257$ nm (bottom, c-d),
        differing only in  $\xi_L=51.4, 25.7, 12.85,$ and 6.425 nm.
        illuminated by a Gaussian beam with $\theta_0=-10^{\circ}$,
        $\lambda=629.9$ nm, and $W=L/4\cos\theta_0$. Left (a and c):
        $s$ polarization; right (b and d): $p$ polarization.}
\label{fig_ff}
\end{figure}

First, we present in Fig.~\ref{fig_ff} the angular distributions of
the speckle pattern intensities scattered from self-affine Ag surface
profiles with identical properties ($D=1.9$, $L=5.14\,\mu$m, and
$\delta$) except for $\xi_L=51.4, 25.7, 12.85,$ and 6.425 nm, as
mentioned in the preceding section.  Both $s$- and $p$-polarized
incident ($\theta_0=-10^{\circ}$) Gaussian beams are considered with
wavelength $\lambda=2\pi c/\omega= 629.9$ nm and
$W=(L/4)\cos\theta_0$, the dielectric constant of Ag at this frequency
being\cite{palik} $\epsilon=-15+\imath$.  The upper plots correspond
to moderately rough surfaces with $\delta=51.4$ nm and topothesy
$l=24$ nm (the corresponding surface profiles will be shown in
Fig.~\ref{fig_sf_1_s}d), beyond the reach of standard approximate
treatments such as the KA or perturbation theories. No significant
differences are found (note the semi-log scale in order to enhance
them) for $s$ polarization, nor even away from the bright specular
peak. In the case of $p$ polarization, a weak decrease in the specular
peak with decreasing $\xi_L$ is barely observed, whereas small
differences appear in the speckle pattern away from the specular peak.

We now consider rougher surfaces with $\delta=257$ nm and topothesy
$l=143$ nm (see Figs.~\ref{fig_ff}c and~\ref{fig_ff}d). Appreciable
changes are observed in the $s$-polarized speckle with decreasing
$\xi_L$, although the patterns are qualitatively similar. However, in
the case of $p$-polarized speckle patterns, differences are both
quantitative and qualitatively notorious. Note the absence of specular
peaks in both polarizations. Therefore, in light of
Figs.~\ref{fig_ff}a and~\ref{fig_ff}d, it is seen that the details at
nanometer scales produce no change in the $s$-polarized speckle
patterns for moderately rough self-affine profiles, appearing only
small quantitative variations for very large rms height. For $p$
polarization, nonetheless, small changes are already observed for
moderately rough profiles, becoming dramatic for large roughness.

\begin{figure}[h]
\includegraphics[width=\columnwidth]{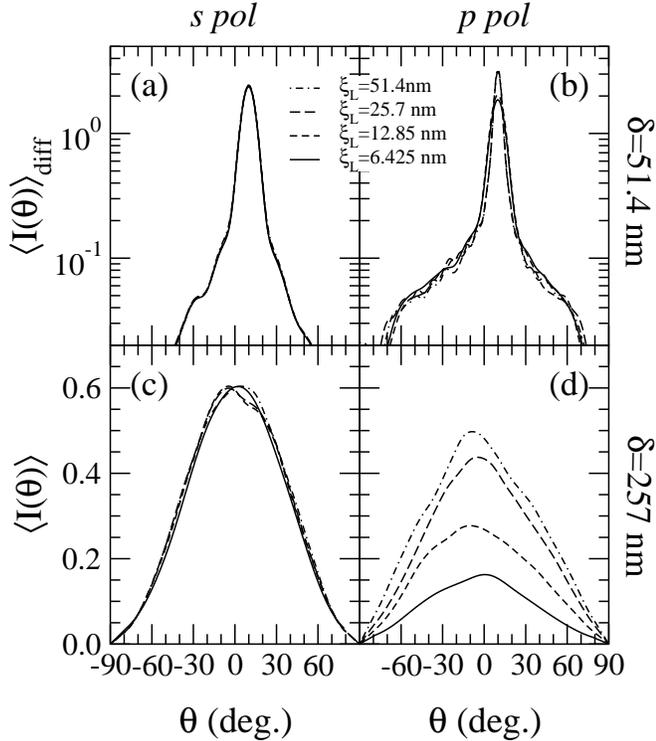}
\caption{Same as in Fig.~\protect{\ref{fig_ff}} but for the mean scattered
        intensity averaged over $N_r=200$ realizations.}
\label{fig_aff}
\end{figure}

Next, the influence of nanoscales on the mean scattered intensities is
analyzed. This is done in Figs.~\ref{fig_aff} after averaging speckle
pattern distributions over an ensemble of $N_{rea}=200$ realizations.
Experimentally, this can be achieved either by also ensemble
averaging, or by illuminating a very large surface and collecting a
large amount of speckles at every scattering angle (or a combination
of both). In the case of the self-affine fractals with $\delta=51.4$
nm (see Figs.~\ref{fig_aff}a and~\ref{fig_aff}b), the diffuse
component of the mean scattered intensity in $s$ polarization is not
altered at all with decreasing lower scale cutoff. As for the
specular, which amounts to $\sim 32\%$, and not shown in
Fig.~\ref{fig_aff}a), the same is true.  The $p$-polarized diffuse
component in Fig.~\ref{fig_aff}b shows no significant variation at
angles away from the specular direction, whereas a very small, but
observable, decrease in the specular direction is observed when the
lower scale cutoff diminishes. On the other hand, the $p$-polarized
specular component, not shown in Fig.~\ref{fig_aff}b, exhibits a
larger decrement (from $\sim 32\%$ down to $\sim 24\%$) with
diminishing $\xi_L$.  Interestingly, despite the specular components
are not included in Figs.~\ref{fig_aff}a and~\ref{fig_aff}b, the
resulting angular distributions of diffusely scattered intensities
present pronounced peaks at the specular directions. These diffuse
specular peaks have been already discussed in Ref. \onlinecite{wrm97} for
self-affine fractals with various fractal dimensions but a higher
lower scale cutoff; the results in Figs.~\ref{fig_aff}a
and~\ref{fig_aff}b demonstrate that such peaks are not substantially
affected by the fractal scales below a hundred nanometers.

\begin{figure}[h]
\includegraphics[width=\columnwidth]{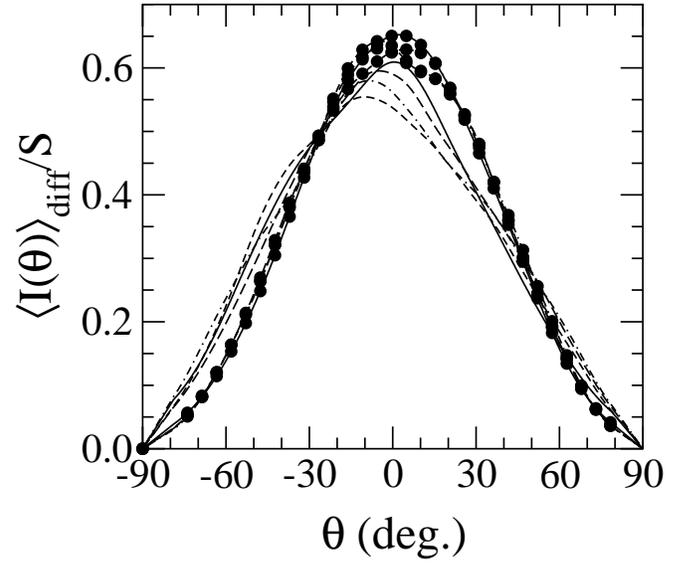}
\caption{Same as in Figs.~\protect{\ref{fig_aff}}c and 
        \protect{\ref{fig_aff}}d but renormalizing 
        by the total reflectance $S$.}
\label{fig_aff_ren}
\end{figure}

Ensemble averaging of the speckle pattern scattered from very rough
($\delta=257$ nm) self-affine fractal surfaces also tends to wash out
differences arising from nanoscale fractal details (see
Figs.~\ref{fig_aff}c and~\ref{fig_aff}d). In fact, no differences are
found in the angular distribution of $s$-polarized, mean scattered
intensities (leaving aside some spurious speckle noise due to the
limited number of realizations employed), even though quantitative
changes have been found above in the speckle patterns for single
realizations (see Fig.~\ref{fig_ff}c). Furthermore, the remarkable
qualitative and quantitative changes in the $p$-polarized speckle
spots seen in Fig.~\ref{fig_ff}d reduce to significant, though only
quantitative, variations upon ensemble averaging as shown in
Fig.~\ref{fig_aff}d. And if the effect of absorption, defined as
$A=1-S$ ($S$ being the normalized reflectance) is taken into account,
the resulting renormalized angular distributions of $p$-polarized mean
scattered intensities, $\langle I(\theta)\rangle /S$, exhibit little
differences with decreasing fractal lower scale cutoff (see
Fig.~\ref{fig_aff_ren}). Incidentally, note that the remaining speckle
noise in Figs.~\ref{fig_aff}d and~\ref{fig_aff_ren} hinders the
possible observation of a backscattering peak due to the multiple
scattering of SPP.\cite{wrm97}

\section{NEAR FIELD}
\label{sec_nf}

We now turn to the investigation of the influence on the near EM field
of the lower scale cutoffs of the self-affine fractal surfaces whose
far-field scattering properties have been studied in the preceding
section. Near field intensity distributions are relevant for the
information they provide on the scattering process, and can be
measured through near-field optical microscopy.

\begin{figure}[h]
\includegraphics[width=\columnwidth]{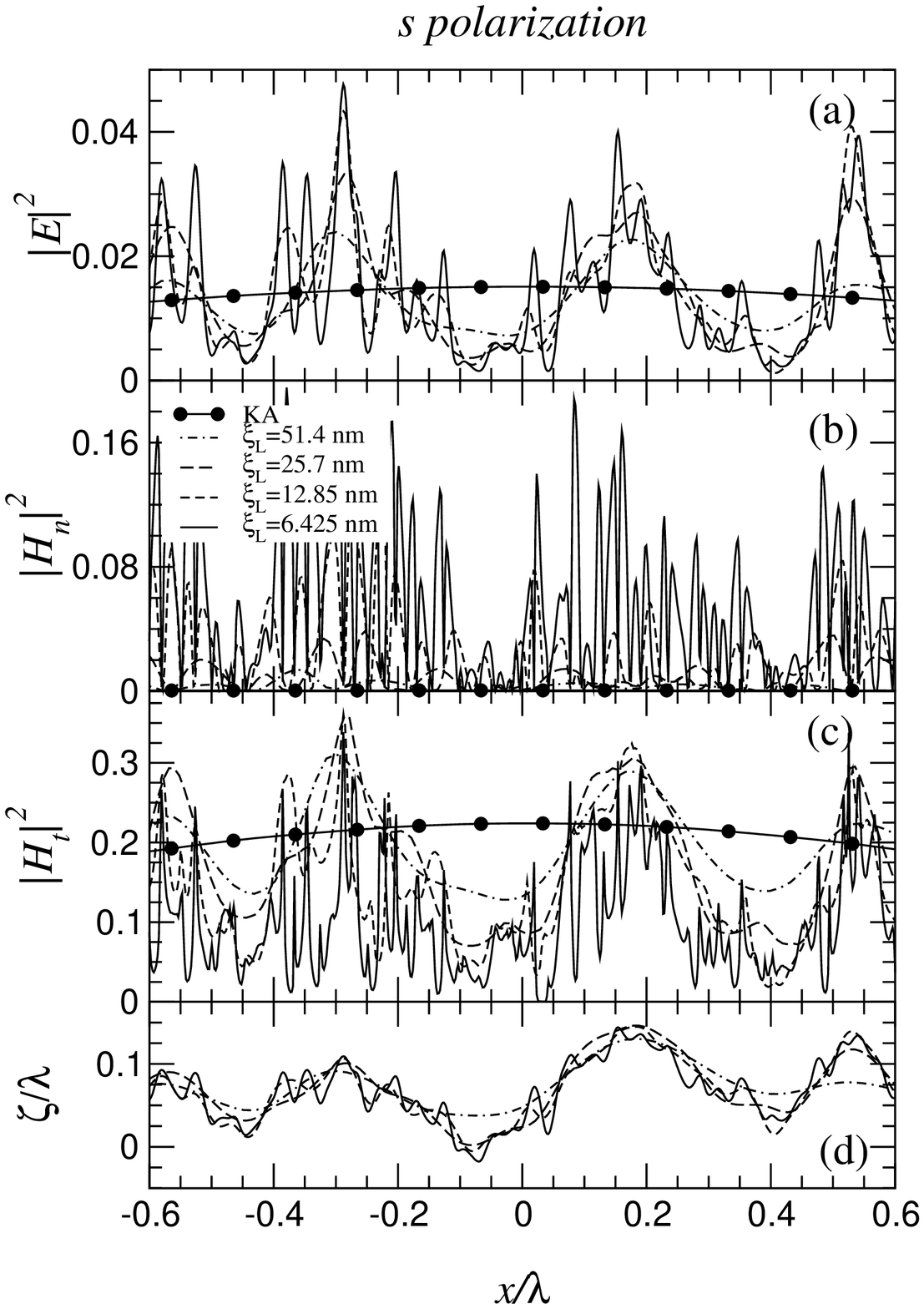}
\caption{Surface electric (a) and magnetic, normal (b) and tangential
        (c), field intensities on the center spot of the illuminated
        area $L=5.14\,\mu$m for $s$-polarized scattering with
        $\theta_0=-10^{\circ}$, $\lambda=629.9$ nm, and
        $W=L/4\cos\theta_0$, from Ag fractal surfaces with $D=1.9$,
        $\delta=51.4$ nm, and $\xi_L=51.4, 25.7, 12.85,$ and 6.425 nm.
        The KA field intensity is also included (see text).
        (d) The corresponding surface profiles. }
\label{fig_sf_1_s}
\end{figure}

\begin{figure}[h]
\includegraphics[width=\columnwidth]{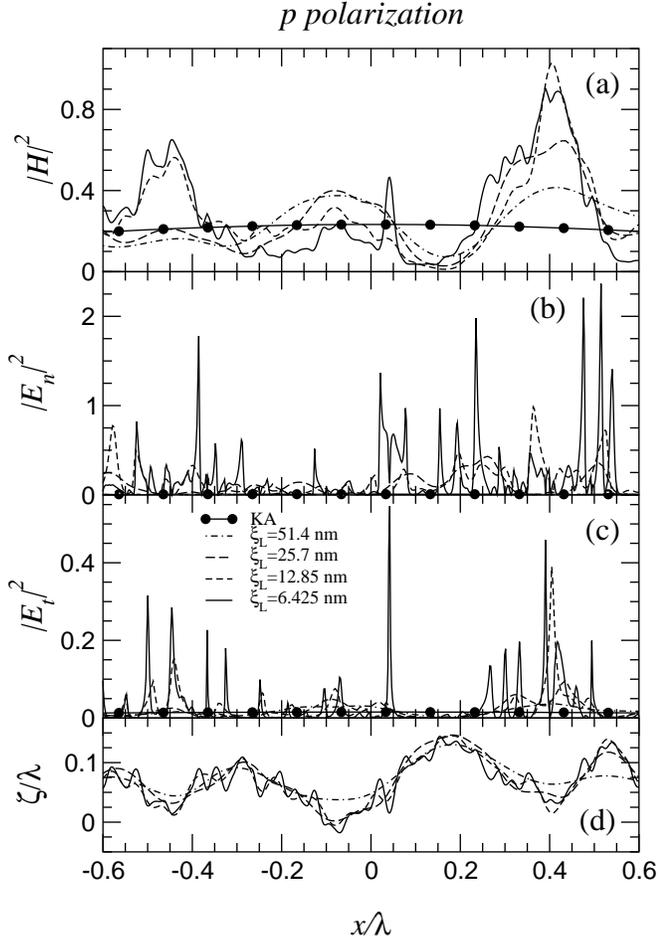}
\caption{Surface magnetic (a) and electric, normal (b) and tangential
        (c), field intensities on the center spot of the illuminated
        area $L=5.14\,\mu$m for $p$-polarized scattering with
        $\theta_0=-10^{\circ}$, $\lambda=629.9$ nm, and
        $W=L/4\cos\theta_0$, from Ag fractal surfaces with $D=1.9$,
        $\delta=51.4$ nm, and $\xi_L=51.4, 25.7, 12.85,$ and 6.425 nm.
        The KA field intensity is also included (see text).
        (d) The corresponding surface profiles. }
\label{fig_sf_1_p}
\end{figure}

First, we calculate the intensities of all the EM field components on
the surface. The results for the moderately rough fractals with
$\delta=51.4$ nm employed in the speckle pattern calculations in
Fig.~\ref{fig_ff} are shown in Figs.~\ref{fig_sf_1_s} 
and~\ref{fig_sf_1_p} for $s$ and $p$ polarization, respectively.
The corresponding surface profiles are shown in the bottom,
Figs.~\ref{fig_sf_1_s}d and Figs.~\ref{fig_sf_1_p}d.  The nonzero
components of the EM field being plotted are (cf.
Sec.~\ref{sec_eq}): the tangential, perpendicular to the incident
plane, electric (respectively, magnetic) field and the normal and
tangential (in the plane of incidence) magnetic (respectively,
electric) field, in the case of $s$ (respectively, $p$) polarization.
For the sake of clarity, only the central part of the illuminated
surface is shown.

Before analyzing the spatial distributions, let us
recall what is expected within the KA, namely, the sum of the incident
and the specularly (locally) reflected fields. Since the angular
variation of the Fresnel coefficients for metals at this frequency is
small, the local variations of the specular field can be
neglected. Therefore, the KA field intensities are approximately given
by those for a planar surface, namely:
\begin{subequations}\begin{eqnarray}
\mid\!E^{(s,KA)}\!\mid^2\approx &&
\mid\! 1+{\cal R}_s\!\mid^2 \mid\! E^{(s,i)}\!\mid^2 \\
\mid\!H_n^{(s,KA)}\!\mid^2\approx && \sin^2\theta_0
\mid\! 1+{\cal R}_s\!\mid^2\mid\! H^{(s,i)}\!\mid  \\
\mid\!H_t^{(s,KA)}\!\mid^2\approx && \cos^2\theta_0
\mid\! 1-{\cal R}_s\!\mid^2\mid\! H^{(s,i)}\!\mid ,
\end{eqnarray}\label{eq_feka_s}\end{subequations}
for $s$ polarization; and
\begin{subequations}\begin{eqnarray}
\mid\!H^{(p,KA)}\!\mid^2\approx &&
\mid\! 1+{\cal R}_p\!\mid^2\mid\! H^{(p,i)}\!\mid^2 \\
\mid\!E_n^{(p,KA)}\!\mid^2\approx && \sin^2\theta_0
\mid\! 1+{\cal R}_p\!\mid^2 \mid\! E^{(p,i)}\!\mid  \\
\mid\!E_t^{(p,KA)}\!\mid^2\approx && \cos^2\theta_0
\mid\! 1-{\cal R}_p\!\mid^2\mid\! E^{(p,i)}\!\mid ,
\end{eqnarray}\label{eq_feka_p}\end{subequations}
for $p$ polarization, ${\cal R}_s$ and ${\cal R}_p$ being the
corresponding Fresnel coefficients for $\theta_0$ on a planar metal
surface. The latter field intensities are plotted in
Figs.~\ref{fig_sf_1_s} and~\ref{fig_sf_1_p}. As expected, the KA does
not hold even for the moderately rough surface profiles used therein,
nor does perturbation theory (recall that the planar surface field
intensities can be considered the zeroth order approximation in the
small-amplitude perturbation expansion of the field), but they provide
the background about which the actual surface EM field intensities
strongly vary.

In the case of $s$ polarization, Fig.~\ref{fig_sf_1_s}, significant
variations appear in the surface EM field upon decreasing the lower
scale cutoff. No appreciable differences have been found
in the corresponding far field speckle patterns in Fig.~\ref{fig_ff}a
and~\ref{fig_ff}b, so that the surface field nanoscale fluctuations
are mostly due to the evanescent components. Fluctuations about the
background distributions, Eqs. (\ref{eq_feka_s}), become narrower and steeper
the smaller is $\xi_L$, namely, the smaller are the surface
features. This effect is more pronounced for the normal magnetic
field, for which the expected KA background is indeed very
small. Interestingly, it should be noted that the electric field
intensity resembles with positive contrast the surface profile (the
same is true for the total, tangential plus normal, magnetic field
intensity), which is not the case in general.\cite{remi95}
Although not valid from a quantitative standpoint, small
amplitude perturbation theory can provide a qualitative
explanation. The first-order term of the surface electric field in
powers of the surface height, which in turn gives the first-order
correction to the planar surface background,\cite{remi95,jj95} can be
cast for this particular scattering geometry, polarization and near
normal incidence in the form of a convolution integral involving the
surface profile function. We have verified, though not shown here,
that by simply increasing the angle of incidence up to
$\theta_0=40^{\circ}$ the resemblance is slightly lost.

Upon illuminating with $p$-polarized light, see Fig.~\ref{fig_sf_1_p},
the surface EM field intensity distributions become more complicated,
with larger variations from one realization to another with
diminishing $\xi_L$, and no resemblance whatsoever with the surface
profiles. These variations are considerably larger than those for the
far-field speckle patterns, indicating the crucial role played by the
evanescent components. Furthermore, the excitation of SPP propagating
along the surface and reradiating into vacuum mediates the scattering
process.  In fact, the spatial frequency of the large oscillations
about the background in the surface EM field intensity is related to
the SPP wavelength;\cite{jcp98} this is more easily observed in the
surface magnetic field in Fig.~\ref{fig_sf_1_p}a (also in the total
electric field, not shown here) for the profiles with higher $\xi_L$.
On the other hand, the surface normal electric field component for the
smaller $\xi_L$ (see Fig.~\ref{fig_sf_1_p}b) reveals the appearance of
narrow and bright spots where an enhancement of the field intensity 
of nearly two orders of magnitude
occur, despite the relatively low value of $\delta$. This clearly
manifests the crucial role played by the lower scale cutoff in the
excitation of localized optical modes.\cite{ol01}

\begin{figure}[h]
\includegraphics[width=\columnwidth]{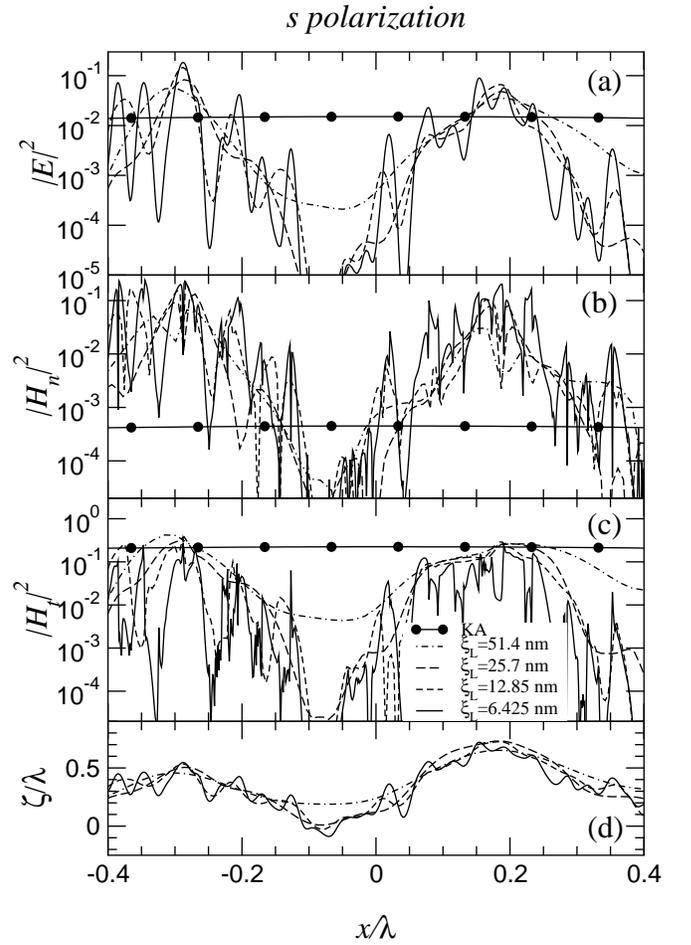}
\caption{Surface EM field intensities (in a semi-$\log_{10}$ scale)
        for $s$ polarization as in Fig.~\protect{\ref{fig_sf_1_s}},
        but for rougher surface profiles (d) with  $\delta=257$ nm.  }
\label{fig_sf_5_s}
\end{figure}

\begin{figure}[h]
\includegraphics[width=\columnwidth]{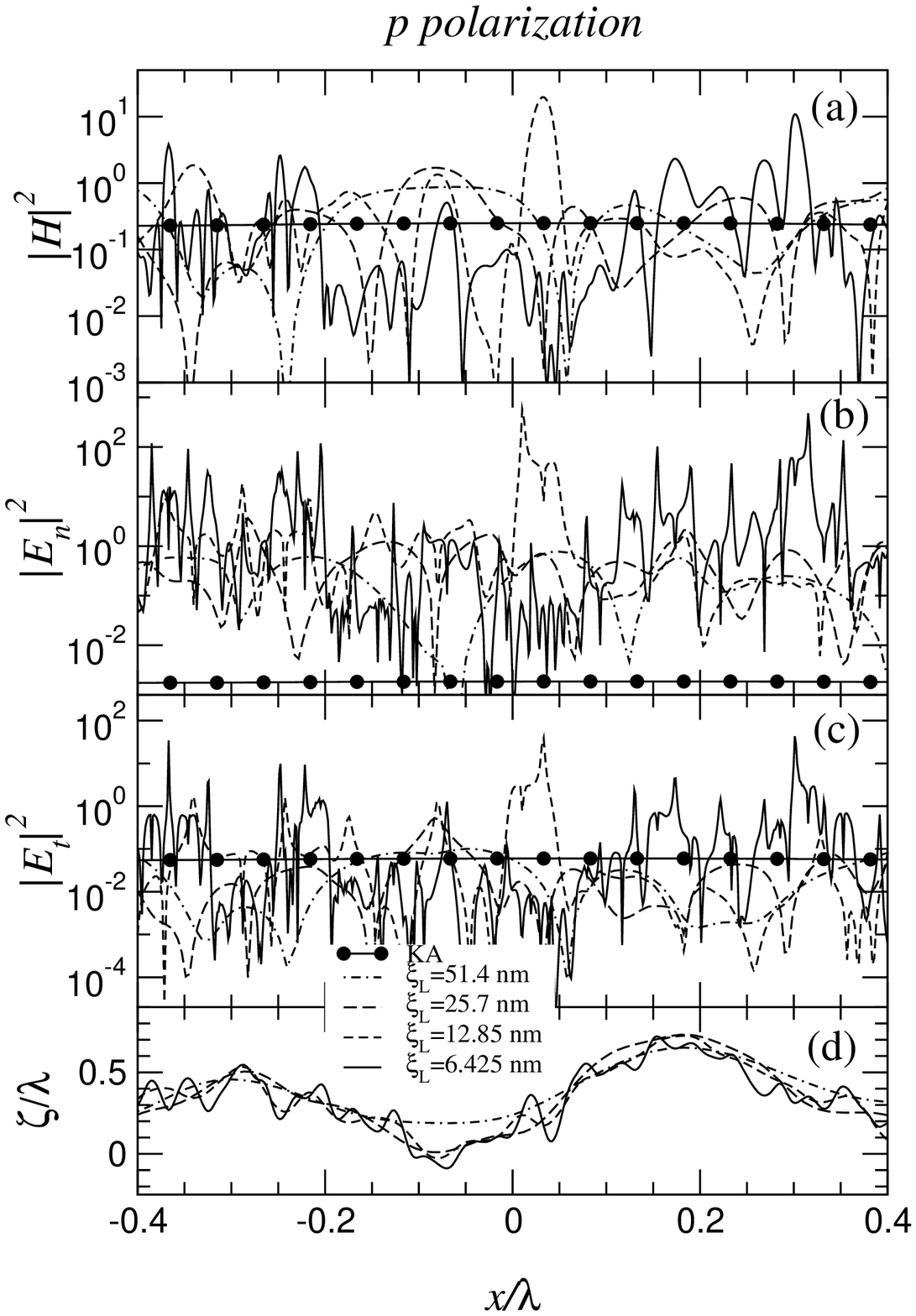}
\caption{Surface EM field intensities (in a semi-$\log_{10}$ scale)
        for $p$ polarization as in Fig.~\protect{\ref{fig_sf_1_p}},
        but for rougher surface profiles (d) with  $\delta=257$ nm.  }
\label{fig_sf_5_p}
\end{figure}

What if the surface roughness parameter is increased? In
Figs.~\ref{fig_sf_5_s} and~\ref{fig_sf_5_p}, the surface EM field
intensities are plotted for $s$ and $p$ polarization, respectively, as
in Figs.~\ref{fig_sf_1_s} and~\ref{fig_sf_1_p}, but for $\delta=257$
nm.  Due to the large differences introduced by $\xi_L$, a semi-log
scale is used and only a small surface region about one wavelength
long is shown. In the case of $s$ polarization, the differences in the
surface EM field intensities from one realization to another with
lower $\xi_L$ are remarkable, and quantitatively larger than those
differences shown above for smoother surface profiles; the qualitative
behavior is however similar.  Lower-$\xi_L$ surfaces lead to surface
EM field intensities with higher spatial frequency components, but the
smoothly varying envelopes correspondig to smaller spatial frequencies
are largely preserved. The former are mostly evanescent components
that contribute little to the far-field intensities, which are
basically originated in the propagating components, and thus quite
similar for all surface profiles (see Fig.~\ref{fig_ff}c). It should
be remarked that no localized optical excitations seem to appear, as
expected since no SPP can be excited, nor significant EM field
enhancements are found.  Note that, thanks to the semi-logarithmic
scale used in Figs.~\ref{fig_sf_5_s} for the surface field
intensities, a weak qualitative resemblance of positive contrast with
the surface profiles can be seen.

The $p$-polarized surface EM field intensities presented in
Fig.~\ref{fig_sf_5_p} reveal very strong changes in all field
components with lower $\xi_L$.  These changes are considerably
stronger than those for the smoother surfaces (see
Fig.~\ref{fig_sf_1_p}).  As mentioned above, narrow, large peaks tend
to concentrate at bright spots in the form of {\it localized optical
modes}, exhibiting very large electric field enhancements
preferentially polarized along the normal to the surface, in agreement
with the SPP polarization, although significantly large enhancements
of the tangential electric field intensity are also found. In
addition, large magnetic field intensities abound, not necessarily at
the same electric field bright spots. The polarization of the modes
has been discussed in Ref. \onlinecite{prb00} for rougher self-affine
fractals with large, but yet subwavelength, $\xi_L$.  It should be
emphasized that the nanoscale lower cutoff largely contributes to the
building and strengthening of such optical excitations.

\begin{figure}[h]
\includegraphics[width=\columnwidth]{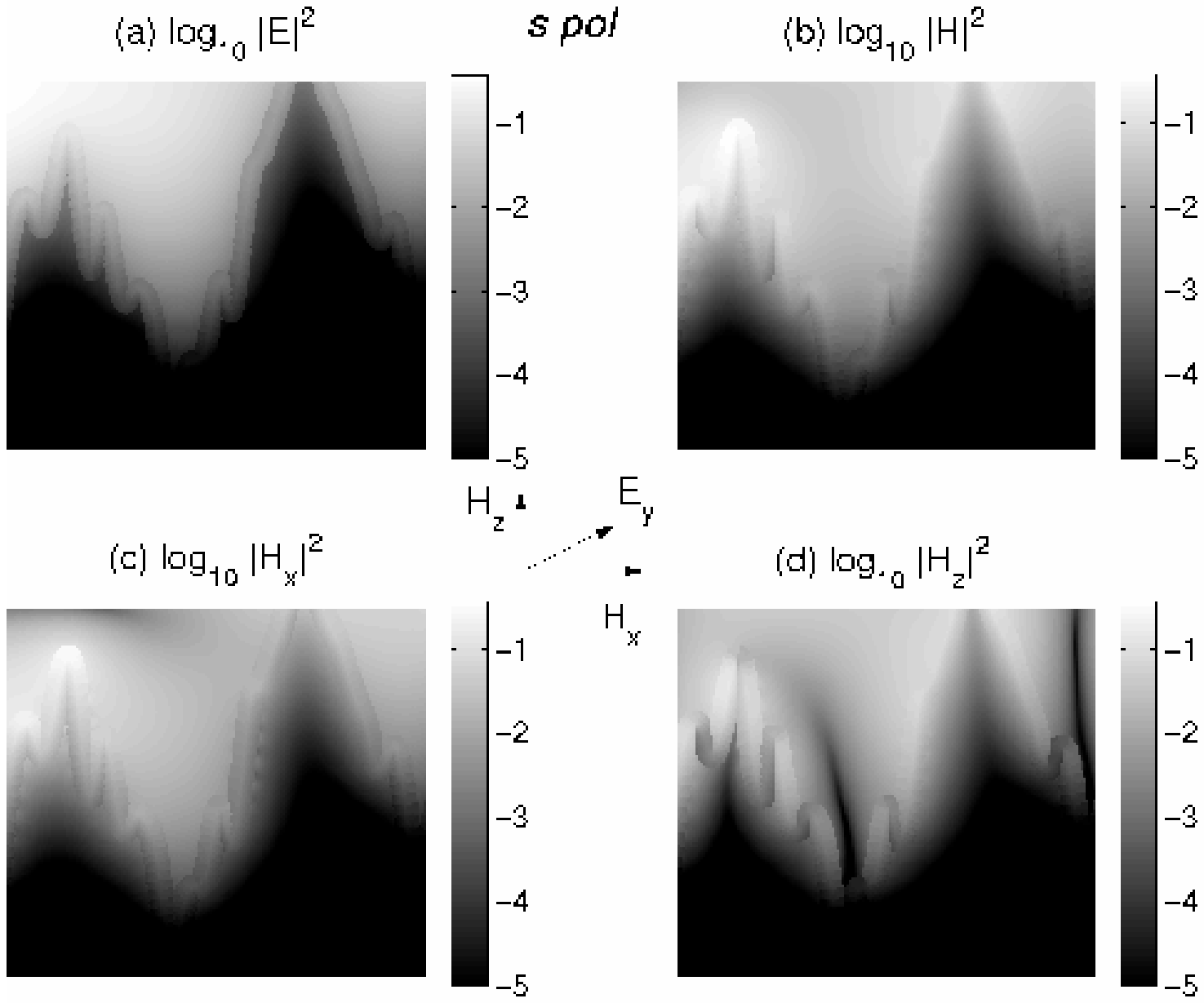}
\caption{Near field intensity images in $s$ polarization
        [(a) electric, (b) magnetic, the
        former split into (c) $x$, horizontal component, and (d) $z$,
        vertical component, all of them in a $\log_{10}$ scale] in
        an area of 0.5$\times 0.5$ $\mu$m$^2$ close to the fractal
        surface in Fig.~\protect{\ref{fig_sf_5_s}}d with $\delta=257$ nm
        and $\xi_L=12.85$ nm. Other parameters as in
        Fig.~\protect{\ref{fig_sf_5_s}}.}
\label{fig_nf_s}
\end{figure}

\begin{figure}[h]
\includegraphics[width=\columnwidth]{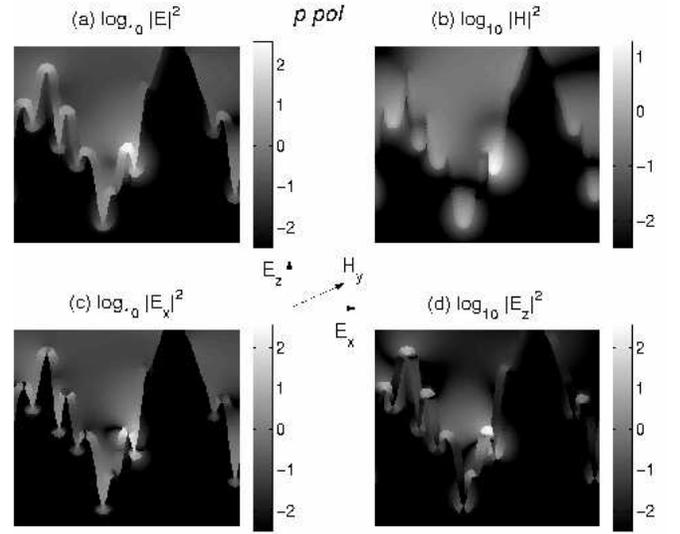}
\caption{Near field intensity images in $p$ polarization 
        [(a) electric, (b) magnetic, the
        latter split into (c) $x$, horizontal component, and (d) $z$,
        vertical component, all of them in a $\log_{10}$ scale] in
        an area of 0.5$\times 0.5$ $\mu$m$^2$ close to the fractal
        surface in Fig.~\protect{\ref{fig_sf_5_p}}d with $\delta=257$ nm
        and $\xi_L=12.85$ nm, where a strong {\it localized optical
        mode} is observed. Other parameters as in
        Fig.~\protect{\ref{fig_sf_5_p}}.}
\label{fig_nf_p}
\end{figure}

Let us plot the EM field in the near vicinity of the rough interface.
This is done in Figs.~\ref{fig_nf_s} and ~\ref{fig_nf_p}, where the
near-field intensity maps in a logarithmic scale for $s$ and $p$
polarization, respectively, are shown in a region around the origin of
the surface profile with $\xi_L=12.85$ nm in Figs.~\ref{fig_sf_5_s}d
and~\ref{fig_sf_5_p}d.  All the EM field components are included for
the sake of completeness; recall that it has been recently reported
that the magnetic field intensity can be also probed through PSTM for
certain experimental configurations.\cite{dev00} The actual surface
profile, though not explicitly depicted, can be inferred from the
fairly black, metal regions due to the evanescent behavior of the EM
fields inside metal with a small skin depth ($d\approx 25$ nm).  

Note that no bright spot whatsoever is observed in the EM field
intensity maps in the case of $s$ polarization, Fig.~\ref{fig_nf_s},
which exhibits no relevant features in the near field. The actual
interface appears slightly blurred, since the minima have been reduced
to enhance contrast. Incidentally, the continuity of the (tangential)
electric field is satisfied in Fig.~\ref{fig_nf_s}a, as well as the
(dis)continuity of the (normal) tangential magnetic field in
Figs.~\ref{fig_nf_s}c and \ref{fig_nf_s}d.

On the other hand, a very bright spot is found at the local maximum in
the central part of the intensity map in Fig.~\ref{fig_nf_p}a: The
intensity field enhancement at such spot is $\mid\!
E\!\mid^2/\mid\!E^{(i)}\!\mid^2\sim 10^4$ (recall that the maximum
scale of EM field intensities in Fig.~\ref{fig_nf_s} is about 3 orders
of magnitude lower than those in Fig.~\ref{fig_nf_p}). Note that there
seems to be a bright, though weaker, magnetic spot associated with the
optical mode (see Fig.~\ref{fig_nf_p}b), which is nonetheless slightly
shifted to the right of the electric field maximum, on the surface
local minimum nearby.  The electric field intensity decays rapidly
upon moving into vacuum away from the bright spot, faster than
expected for the evanescent decay of SPP propagating on a plane. This
may help to explain why localized optical modes experimentally
observed through PSTM yield considerably smaller enhancement
factors,\cite{bozh98} leaving aside the fact that no direct comparison
with theoretical calculations for the actual experimental surface
profile are available.  With regard to the electric field orientation
on the bright spot, Figs.~\ref{fig_nf_p}c and ~\ref{fig_nf_p}d
indicate that the normal electric field component is responsible for
the bright spot, in agreement with Figs. \ref{fig_sf_5_p}b and
\ref{fig_sf_5_p}c; this component corresponds to $\mid\! E_z\!\mid^2$
in regions near surface maxima and minima (see Fig.~\ref{fig_nf_p}d)
and to $\mid\!  E_x\!\mid^2$ near vertical surface walls (see
Fig.~\ref{fig_nf_p}c). It can be also observed that, as expected, the
normal electric field is discontinuous across the interface, the
tangential component being continuous. The continuity of the
(tangential) magnetic field for $p$ polarization is evident in
Fig.~\ref{fig_nf_p}a.

The statistics of the surface electric field enhancement has been
studied elsewhere,\cite{ol01} confirming a significant increase with
decreasing $\xi_L$. It should be remarked that such large surface
electric fields plays a decisive role in SERS and other surface
non-linear optical processes;\cite{sharep,jcp98,shg} in particular,
local values of $\sim 10^4$ can help to explain the existence of
bright spots at which SERS single-molecule detection has been
claimed.\cite{nie97,kneipp97,xu99}

\section{CONCLUDING REMARKS}
\label{sec_con}

To summarize, we have studied the influence of the nanoscale (in the
region below a hundred nanometers) lower cutoff $\xi_L$ on the
scattering of light from one-dimensional, self-affine fractal Ag
surfaces with large fractal dimension $D=1.9$ and for both moderately
($\delta=$ 51.45 nm) and large ($\delta=$ 257 nm) surface height
deviations. Since no approximate methods are applicable for such
roughness parameters, numerical calculations have been carried out on
the basis of the Green's theorem integral equation formulations,
extended to account for all EM field components in the near field
region.  Both far-field and near-field distributions have been
obtained for $s$ and $p$ polarization (no depolarization takes place
in this scattering geometry). We have presented far-field intensity
distributions corresponding to single realizations (speckle patterns)
and averages over an ensemble of realizations.

It has been shown that the details corresponding to scales below a
hundred nanometers in moderately rough self-affine profiles produce no
change in the far field speckle pattern, appearing only small
quantitative variations for very large rms height. For $p$
polarization, nonetheless, small changes are already observed for
moderately rough profiles, becoming dramatic for large
roughness. Ensemble averaging tends to wash out the differences. In
the case of $s$ polarization, the angular distributions of mean
scattered intensities are practically identical, not only for
$\delta=$ 51.45 nm, but also for $\delta=$ 257 nm. For $p$
polarization, the differences encountered in the speckle patterns are
largely suppressed in the mean scattered intensity distributions; the
only effect of the varying nanoscale cutoff $\xi_L$ manifests in the
total reflectance (which is smaller for decreasing $\xi_L$ due to
larger absorption).

Nevertheless, nanoscale features have a strong impact on the near EM
field distributions due to the relevant role played by the evanescent
components. The $s$-polarized surface EM field intensity exhibits
oscillations with higher frequency the smaller $\xi_L$ is. The
amplitude of such oscillations, and thus the changes from one profile
to another with different $\xi_L$, increases with the rms height
$\delta$. No significant electric field enhancements are found in this
polarization. Incidentally, the surface and near electric and magnetic
field intensities vaguely resemble the surface profile with positive
contrast despite the large surface roughness of the profiles. In the
case of $p$ polarization, drastic changes with decreasing $\xi_L$ are
found in the surface EM field stemming from the roughness-induced
excitation of SPP. Large, narrow peaks ({\it localized optical modes})
tend to appear with either increasing rms height or decreasing
nanoscale, predominantly enhancing the normal electric field intensity
(as expected from the SPP electric field orientation), and rapidly
decaying into both vacuum and metal. Our near field intensity maps
around optical excitations can help to interpret PSTM experimental
results. In addition, the large field enhancements associated with
such {\it localized optical modes} on self-affine fractal metal
surfaces can shed light onto SERS and surface nonlinear optical
phenomena.

\begin{acknowledgments}

This work was supported by the Spanish Direcci{\'o}n General de
Investigaci{\'o}n (grant BFM2000-0806) and Comunidad de Madrid
(grant 07M/0111/2000).

\end{acknowledgments}


\end{document}